\documentstyle[12pt]{article}

\newcommand{\be}{\begin{equation}}
\newcommand{\ee}{\end{equation}}
\newcommand{\ks}{k\!\!\!/}
\newcommand{\qs}{q\!\!\!/}
\newcommand{\ppl}{p\!\!/}
\newcommand{\epsilons}{\epsilon\!\!/}

\date{}
\title{
\hfill {\large\rm SLAC-PUB-7257}\\
\hfill {\large\rm September 1996}\vspace*{2cm}\\
The Drell-Yan Process and Factorization \\
in Impact Parameter Space
\thanks{Work partially supported by the Department of Energy, contract
DE-AC03-76SF00515.}\\*[1.5cm]}
\author{S. J.~Brodsky, A.~Hebecker, and E.~Quack\\
{\normalsize\it Stanford Linear Accelerator Center, Stanford University,
Stanford, CA 94309}}

\input psfig
\begin{document}

\setlength{\baselineskip}{18pt}
\maketitle
\begin{abstract}
\noindent
The cross section for Drell-Yan pair production in the limit of small 
$x_{\mbox{\scriptsize{target}}}$ is derived in the rest frame of the target 
hadron. Our calculation is based on the fundamental quantity 
$\sigma(\rho)$, the cross section for the scattering of a $q\bar{q}$-pair  
with fixed transverse separation $\rho$ off a hadronic target. As in deep 
inelastic scattering the result can be given in terms of integrals of 
$\sigma(\rho)$. This is consistent with well known factorization theorems 
and also relates higher-twist terms in both processes. An analysis of the 
angular distribution of the produced lepton shows that additional 
integrals of $\sigma(\rho)$ can be obtained in the Drell-Yan process, which 
are not measurable in inclusive deep inelastic scattering. 
\end{abstract}
\vspace*{.5cm}
\begin{center}(To be submitted for publication)\end{center}
\thispagestyle{empty}
\newpage

\section{Introduction}
The Drell-Yan (DY) process, i.e. the production of massive lepton pairs in 
hadronic collisions, has remained, together with deep inelastic scattering 
(DIS), one of the most prominent processes in strong interaction physics. 
In recent years, in connection with the availability of high energy 
machines like HERA and the Tevatron, much attention has been devoted to the 
small-$x$ region of QCD, where parton densities become high and 
perturbative methods reach their limits. 

Although extensive work exists in the field of small-$x$ DIS and related 
processes, the small-$x$ limit of lepton pair production has received only 
limited theoretical attention. In our opinion the small-$x$ or high energy 
region of the DY process, i.e. the region where the lepton pair mass $M$ is 
much smaller than the available energy $\sqrt{s}$, deserves study for at 
least two reasons: 

First, it is of general theoretical interest to understand the 
interrelations between the high energy limits of DIS and DY pair 
production on both nucleon and nuclear targets. Though general 
factorization theorems are established (see e.g. \cite{css}), it is still 
worthwhile to develop an intuition for the way they are realized 
specifically in the small-$x$ region. 

Second, the DY process may provide new tools for the experimental 
investigation of the small-$x$ dynamics in QCD. In particular, lepton pair 
production in the region $M^2\ll s$ may be one of the cleanest processes 
for the study of new phenomena in heavy ion collisions at future colliders. 

Recently, a new approach to the DY process has been suggested by 
Kopeliovich \cite{kop}, with the aim to understand the observed nuclear 
shadowing at small $x_{\mbox{\scriptsize{target}}}$ \cite{ald}. It has been 
observed, that in analogy to DIS, the DY cross section at high energies can 
be expressed in terms of the scattering cross section of a color-neutral 
$q\bar{q}$-pair.

In the present investigation, we derive the high energy DY cross section 
in the target rest frame. The dominant underlying process is the scattering 
of a parton from the projectile structure function off the target color 
field. This parton radiates a massive photon, which subsequently decays into 
a lepton pair. Our treatment of the interaction of the projectile parton 
with the target hadron makes use of the high energy limit, but it is not 
restricted to the exchange of a finite number of gluons. 

Using the non-perturbative $q\bar{q}$-cross section $\sigma(\rho)$, where 
$\rho$ is the transverse separation of the pair, a parallel description of 
DIS and DY pair production is presented in the rather general framework 
given above. The cross section $\sigma(\rho)$ appears in DIS since the 
incoming photon splits into a $q\bar{q}$-pair, testing the target field at 
two transverse positions \cite{nz}. Similarly, $\sigma(\rho)$ appears in 
the DY process due to the interference of amplitudes in which the fast 
quark of the projectile hits the target at different impact parameters. 

Our main focus is on the role of the photon polarization. The interplay of 
small and large transverse distances, characterized by different values of 
the parameter $\rho$, is compared in DIS and the DY process for transverse 
and longitudinal photons. In addition, the azimuthal angular correlations 
in the DY process provide a new tool for the investigation of 
$\sigma(\rho)$, which is not available in inclusive DIS. 

The paper is organized as follows: After reviewing the impact parameter 
description of DIS in Sect.~\ref{dis}, an analogous calculation of the 
cross section for DY pair production is presented in Sect.~\ref{dy}. In 
Sect.~\ref{dyad} the angular distribution of the produced lepton is given 
in terms of integrals of the $q\bar{q}$-cross section $\sigma(\rho)$. 
Concluding remarks in Sect.~\ref{con} are followed by an appendix, which 
describes the technical details of the calculation.

\section{Deep inelastic scattering in the target rest frame}\label{dis}
A detailed discussion of small $x$ DIS in the target rest frame and in 
impact parameter space has been given in \cite{nz}. The main 
non-perturbative input is the scattering cross section $\sigma(\rho)$ of a 
quark-antiquark pair with fixed transverse separation $\rho$. In the 
present section this approach is briefly reviewed and reformulated in a way 
allowing straightforward generalization to the DY process. 

\begin{figure}[ht]
\begin{center}
\parbox[b]{8cm}{\psfig{width=6cm,file=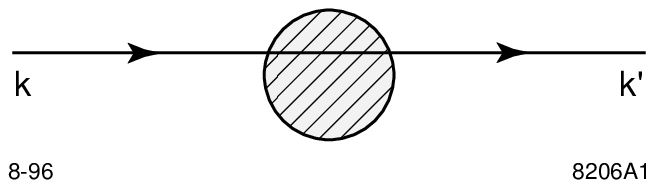}}\\
\end{center}
\refstepcounter{figure}
\label{quark}
{\bf Fig.\ref{quark}} Scattering of a quark off the proton field. 
\end{figure}

Consider first the scattering of a single energetic quark off an external 
color field, e.g. the field of a proton (Fig.~\ref{quark}). The 
complications associated with the color of the quark in the initial and 
final state can be neglected at this point, since the quark amplitude is 
only needed as a building block for the scattering of a color-neutral 
$q\bar{q}$-pair. 

In the high energy limit the soft hadronic field cannot change the energy 
of the quark significantly. Furthermore, we assume helicity conservation 
and linear growth of the amplitude with energy. Therefore, introducing an 
effective vertex $V(k',k)$, the amplitude can be given in the form 
\be
i2\pi\delta(k_0'-k_0)T_{fi}=\bar{u}_{s'}(k')V(k',k)u_s(k)=i2\pi\delta(k_0'
-k_0)\,2k_0\,\delta_{s's}\,\tilde{t}_q(k_\perp'-k_\perp)\, .\label{tdef}
\ee
Here $\tilde{t}_q(p_\perp)$ can be interpreted as the Fourier transform of 
an impact parameter space amplitude, 
\be
\tilde{t}_q(p_\perp)=\int d^2x_\perp\,t_q(x_\perp)\,e^{-ip_\perp x_\perp}
\, .
\ee
Note that $t$ is a matrix in color space. 

If the interaction of the quark with the color field is treated in the 
non-Abelian eikonal approximation, $t_q$ is given explicitly by \cite{nac} 
\be
1+it_q(x_\perp)=F(x_\perp)=P\exp\left(-\frac{i}{2}\int_{-\infty}^{\infty}A_-
(x_+,x_\perp)dx_+\right)\, .\label{eik}
\ee
Here $x_\pm=x_0\pm x_3$ are the light-cone components of $x$, 
$A(x_+,x_\perp)$ is the gauge field, and the path ordering $P$ sets the 
field at smallest $x_+$ to the rightmost position. The $x_-$-dependence of 
$A$ is irrelevant as long as it is sufficiently smooth. 

However, our analysis in the following does not rely on the specific form 
of $t_q$ provided by the eikonal approximation Eq.~(\ref{eik}). 

\begin{figure}[ht]
\begin{center}
\parbox[b]{10cm}{\psfig{width=9.5cm,file=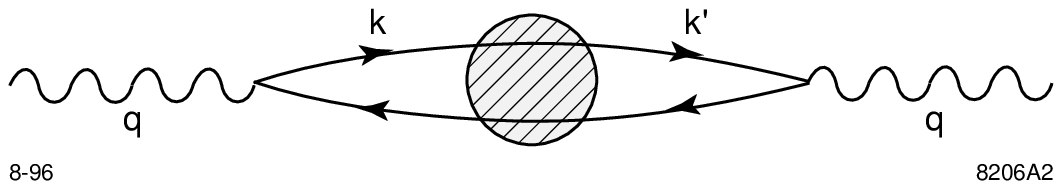}}\\
\end{center}
\refstepcounter{figure}
\label{vp}
{\bf Fig.\ref{vp}} Elastic forward scattering of a virtual photon off the
proton field.
\end{figure}

Consider now the forward elastic scattering of a photon with virtuality 
$Q^2$ off an external field which is related to the total cross section 
via the optical theorem. In the limit of very high photon energy, 
corresponding to the small-$x$ region, the dominant process is the 
fluctuation of the photon into a $q\bar{q}$-pair long before the target 
(see Fig.~\ref{vp}). The quark and antiquark then scatter independently off 
the external field and recombine far behind the target. The virtualities of 
the quarks, which are small compared to their energies, do not affect their 
effective scattering vertices. They enter the calculation only via the 
explicit quark propagators connected to the photon. 

The necessary calculations have been performed many years ago for the 
Abelian case in light-cone quantization \cite{bks} and, more recently, in 
a covariant approach, treating two gluon exchange in the high energy 
limit \cite{nz}.

In the notation of \cite{nz} the transverse and longitudinal photon cross
sections read
\be
\sigma_{T,L}=\int_0^1d\alpha\int d^2\rho_\perp\sigma(\rho)W_{T,L}
(\alpha,\rho)\, ,
\ee
where $\rho=|\rho_\perp|$ is the transverse separation of quark and 
antiquark when they hit the target proton, and $\alpha$ is the longitudinal 
momentum fraction of the photon carried by the quark. The cross section 
$\sigma(\rho)$ for the scattering of the $q\bar{q}$-pair is given by 
\be
\sigma(\rho)=\frac{2}{3}\,\mbox{Im}\int d^2x_\perp\,\mbox{tr}\left[it_q
(x_\perp)t_{\bar{q}}(x_\perp+\rho_\perp)+t_q(x_\perp)+t_{\bar{q}}(x_\perp+
\rho_\perp)\right].\label{sigmaqq}
\ee
Here $t_q(x_\perp)$ is the quark scattering amplitude in impact parameter 
space introduced above and $t_{\bar{q}}(x_\perp+\rho_\perp)$ is its 
antiquark analogue. The last two terms in Eq.~(\ref{sigmaqq}) correspond to 
diagrams where only the quark or only the antiquark is scattered. 

\begin{figure}[ht]
\begin{center}
\parbox[b]{7cm}{\psfig{width=6.5cm,file=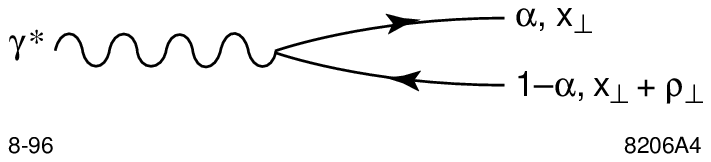}}\\
\end{center}
\refstepcounter{figure}
\label{lcwf}
{\bf Fig.\ref{lcwf}} Light-cone wave function of the virtual photon in the 
mixed representation. 
\end{figure}

We denote by $W_{T,L}$ the squares of the light-cone wave functions of a 
transverse photon and a longitudinal photon with virtuality $Q^2$. In the 
case of one massless quark generation with one unit of electric charge they 
are given by 
\begin{eqnarray}
W_T(\alpha,\rho)&=&\frac{6\alpha_{\mbox{\scriptsize em}}}{(2\pi)^2}N^2
[\alpha^2+(1-\alpha)^2]K_1^2(N\rho)\label{wt}\\ \nonumber\\
W_L(\alpha,\rho)&=&\frac{24\alpha_{\mbox{\scriptsize em}}}{(2\pi)^2}N^2
[\alpha(1-\alpha)]K_0^2(N\rho)\, ,\label{wl}
\end{eqnarray}
where $N^2=N^2(\alpha,Q^2)\equiv\alpha(1-\alpha)Q^2$ and $K_{0,1}$ are 
modified Bessel functions. As is illustrated in Fig.~\ref{lcwf} the 
variables $\alpha$ and $1\!-\!\alpha$ denote the longitudinal momentum 
fractions of the photon carried by quark and antiquark. 

Notice that the color factor $1/3$ in Eq.~(\ref{sigmaqq}) is compensated 
by a factor 3 in Eqs. (\ref{wt}),(\ref{wl}), so that $\sigma(\rho)$ is the 
cross section for one color-neutral $q\bar{q}$-pair and the color summation 
is included in the definition of $W_{T,L}$. 

Consider now the region of a relatively soft quark, $\alpha<\Lambda^2/Q^2$, 
with a hadronic scale $\Lambda\ll Q$. This region, where $N^2\simeq\alpha 
Q^2=a^2$, corresponding to Bjorken's aligned jet model \cite{bj} (see also 
\cite {lps}), gives a higher-twist contribution to $\sigma_L$ and a 
leading-twist contribution to $\sigma_T$, 
\be
\sigma_{T,\bar{q}}=\frac{6\alpha_{\mbox{\scriptsize em}}}{(2\pi)^2Q^2}
\int_0^{\Lambda^2}da^2\int d^2\rho_\perp a^2K_1^2(a\rho)\sigma(\rho)\, .
\ee
A possible interpretation of DIS in this kinematical region is the 
splitting of the photon into a fast, on-shell antiquark and a soft quark, 
which is not far off-shell. In a frame where the proton is fast, the latter 
one corresponds to an incoming antiquark described by a scaling antiquark 
distribution $\bar{q}(x)$. We thus denote the cross section by 
$\sigma_{T,\bar{q}}$. Using the standard formula for the contribution of 
the antiquark structure to the transverse cross section, 
\be
\sigma_{T,\bar{q}}=\frac{(2\pi)^2\alpha_{\mbox{\scriptsize em}}}{Q^2}x
\bar{q}(x)\, ,
\ee
the antiquark distribution can then be given in terms of the 
$q\bar{q}$-cross section, 
\be
x\bar{q}(x)=\frac{6}{(2\pi)^4}\int_0^{\Lambda^2}da^2\int d^2\rho_\perp
a^2K_1^2(a\rho)\sigma(\rho)\, .\label{qdis}
\ee
This formula will be reproduced below from the impact parameter space 
description of DY pair production at small 
$x_{\mbox{\scriptsize{target}}}$, in agreement with factorization. 

The above discussion in terms of scatterings off an external field can be 
generalized to the case of a realistic hadron target by summing 
appropriately over all contributing field configurations. Such an approach 
has already been used for the treatment of DIS in \cite{bal} and for the 
treatment of diffraction in \cite{bdh}.

\section{Drell-Yan Process in the target rest frame}\label{dy}
In this section the DY pair production cross section at small 
$x_{\mbox{\scriptsize{target}}}$ will be calculated in the target rest 
frame. Such an approach has recently been suggested by Kopeliovich in the 
context of nuclear shadowing \cite{kop}. 

Consider the kinematical region where the mass of the produced lepton pair 
is large compared to the hadronic scale, but much smaller than the hadronic 
center of mass energy, $\Lambda^2\ll M^2\ll s$. Furthermore, let the 
longitudinal momentum fraction $x_F$ of the projectile hadron carried by 
the DY pair be large, but not too close to 1. We assume here that the last 
condition allows us to neglect higher-twist contributions from spectator 
partons in the projectile \cite{bb}. 

In the parton model the above process is described as the fusion of a 
projectile quark with momentum fraction $x\approx x_F$ and a target 
antiquark with momentum fraction $x_{\mbox{\scriptsize{target}}}\approx 
M^2/sx_F\ll1$. (Here and below we neglect the antiquark distribution of the 
projectile at the relevant values of $x_F$.) 

\begin{figure}[ht]
\begin{center}
\parbox[b]{12cm}{\psfig{width=11.5cm,file=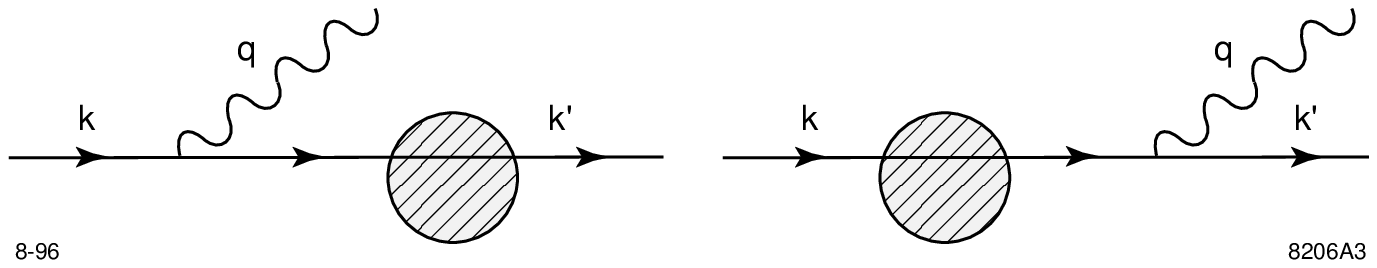}}\\
\end{center}
\refstepcounter{figure}
\label{mp}
{\bf Fig.\ref{mp}} Production of a massive photon by a quark scattering off
the target field. A quark with momentum $k$ interacts with an external 
field producing a photon with momentum $q$ and an outgoing quark with 
momentum $k'$.
\end{figure}

However, a different physical picture of this process is appropriate in the 
target rest frame: A large-$x$ quark of the projectile scatters off 
the gluonic field of the target and radiates a massive photon, which 
subsequently decays into leptons (compare \cite{bln}). The two relevant 
diagrams, corresponding to the photon being radiated before or after the 
interaction with the target, are shown in Fig.~\ref{mp}. Diagrams where the 
quark interacts with the target both before and after the photon vertex are 
suppressed in the high energy limit \cite{bh}. Note that in the above 
approach no antiquark distribution of the target has to be introduced. 
Instead, its effect is produced by the target color field. 

In the high energy limit, i.e. $q_0,k_0,k_0'\gg M^2$, the corresponding 
cross section, including the decay of the photon into the lepton pair, 
reads ($e^2=4\pi\alpha_{\mbox{\scriptsize em}}$)
\be
\frac{d\hat{\sigma}}{dx_FdM^2}=\frac{e^2}{72(2\pi)^3}\cdot\frac{1}
{x_Fk_0k_0'M^2}\int\frac{d^2q_\perp}{(2\pi)^2}\frac{d^2k_\perp'}{(2\pi)^2}
|T|^2\, .\label{dycs}
\ee
Here $T$ is the amplitude for the production of the virtual photon, given 
by the sum of the two diagrams in Fig.~\ref{mp},
\be
i2\pi\delta(q_0\!+\!k_0'\!-\!k_0)T_\lambda=e\bar{u}_{s'}(k')\left[V(k',k\!-
\!q)\frac{i}{\ks-\qs}\epsilons_\lambda(q)+\epsilons_\lambda(q)\frac{i}
{\ks'+\qs}V(k'\!+\!q,k)\right]u_s(k)\, .\label{dyt}
\ee
The matrix $V$ is the effective quark scattering vertex introduced in the 
previous section, and $\epsilon(q)$ is the polarization vector of the 
produced photon, accessible via the lepton angular distribution. Averaging 
over $s$ and summation over $s'$ and $\lambda$ is understood in 
Eq.~(\ref{dycs}). 

When the cross section is explicitly calculated, the quark scattering 
amplitudes $t_q(x_\perp)$ implicit in $V$ combine in a way very similar to 
the case of DIS. Therefore, the final result can be expressed in terms of 
the $q\bar{q}$-cross section introduced in the previous section \cite{kop}. 
This cross section arises from the interference of the two diagrams in 
Fig.~\ref{mp}. To understand the parallelism of the DY process and DIS, 
observe that in the DY cross section the product of two quark amplitudes 
tests the external field at two different transverse positions. In DIS this 
corresponds to the quark-antiquark pair wave function of the virtual 
photon, which tests the external field at two transverse positions as well. 
The details of this calculation are presented in the appendix. 

To make the analogy to DIS more apparent, the cross sections for the 
production of the lepton pair via transversely and longitudinally polarized 
photons are given separately: 
\be
\frac{d\sigma_{T,L}}{dx_FdM^2}=\frac{\alpha_{\mbox{\scriptsize em}}}{9(2\pi)
M^2}\!\!\int\limits_0^{(1-x_F)/x_F}\!\!\!\!\!d\alpha\int d^2\rho_\perp
\frac{q\Big(x_F(1+\alpha)\Big)}{(1+\alpha)^2}
\sigma(\rho)W^{DY}_{T,L}(\alpha,\rho)\, .\label{dyex}
\ee
Here $q(x)$ is the quark distribution of the projectile, $\alpha=k_0'/q_0$ 
is the ratio of energies or longitudinal momenta of outgoing quark and 
photon, and $W^{DY}_{T,L}$ are the analogues of the squares of the photon 
wave functions defined for DIS in the previous section,\footnote{
Our formula for the transverse polarization is similar but not identical to 
the result given in \cite{kop}.} 
\begin{eqnarray}
W^{DY}_T(\alpha,\rho)&=&\frac{12\alpha_{\mbox{\scriptsize em}}}{(2\pi)^2}
N^2[\alpha^2+(1+\alpha)^2]K_1^2(N\rho)\label{wdyt}\\ \nonumber\\
W^{DY}_L(\alpha,\rho)&=&\frac{24\alpha_{\mbox{\scriptsize em}}}{(2\pi)^2}N^2
[\alpha(1+\alpha)]K_0^2(N\rho)\, .\label{wdyl}
\end{eqnarray}
As in the DIS case a subsidiary variable $N^2=\alpha(1+\alpha)M^2$ has
been introduced.

We have defined the polarization of the massive photon in the $u$-channel 
frame. In this frame the photon is at rest and the $z$-axis, defining the 
longitudinal polarization vector, is antiparallel to the momentum of the 
target hadron. Since the polarizations are invariant with respect to boosts 
along the $z$-axis, one could also say that the longitudinal polarization is 
defined by the direction of the photon momentum, in a frame where photon 
and target hadron momenta are antiparallel. This last definition makes it 
obvious that the polarizations in the $u$-channel frame of DY pair 
production are analogous to the standard polarization choice in DIS, 
defining $\sigma_T$ and $\sigma_L$. 

Note, that Eqs. (\ref{wdyt}),(\ref{wdyl}) can be obtained from their 
analogues in DIS, Eqs. (\ref{wt}),(\ref{wl}), by the substitutions 
$Q^2\to M^2$ and $1\!-\!\alpha\,\to\,1\!+\!\alpha$. The last substitution 
reflects the fact that the longitudinal parton momenta in units of the 
photon momentum are $\alpha$ and $1-\alpha$ in DIS, as opposed to $\alpha$ 
and $1+\alpha$ in DY pair production. Since the transverse polarizations 
are summed rather than averaged in the DY process, an additional factor of 
$2$ appears in Eq.~(\ref{wdyt}) as compared to Eq.~(\ref{wt}). 

Consider now the region of a relatively soft outgoing quark, 
$\alpha<\Lambda^2/M^2$, with a hadronic scale $\Lambda\ll M$. In analogy to 
the DIS case, this region gives a higher-twist contribution for 
longitudinal polarization and a leading-twist contribution for transverse 
polarization: 
\be
\frac{d\sigma_{T,\bar{q}}}{dx_FdM^2}=\frac{\alpha_{\mbox{\scriptsize em}}^2
q(x_F)}{6\pi^3M^4}\int_0^{\Lambda^2}da^2\int d^2\rho_\perp a^2K_1^2(a\rho)
\sigma(\rho)\, .\label{sdy}
\ee
Here, assuming a sufficiently smooth behavior of $q(x)$, terms suppressed 
by powers of $\Lambda/M$ have been dropped. 

The above kinematical region corresponds to the contribution from the 
antiquark distribution of the target as calculated in the parton model at 
leading order, 
\be
\frac{d\sigma_{T,\bar{q}}}{dx_FdM^2}=\frac{4\pi
\alpha_{\mbox{\scriptsize em}}^2}{9M^4}q(x_F)\cdot x_t\bar{q}(x_t)\, .
\ee
By comparing this formula with Eq.~(\ref{sdy}), an expression for 
$x\bar{q}(x)$ can be derived which is identical to Eq.~(\ref{qdis}) 
obtained in the case of DIS. This, of course, was to be expected in view of 
the factorization theorems (see e.g. \cite{css}) relating DIS and the DY 
process. 

So far the target hadron has been treated simply as a given external color 
field. As already pointed out in the last section, a more realistic model 
has to include an appropriate summation over all contributing field 
configurations. These field configurations, together with the produced 
lepton pair and the projectile remnant, form the final state of the 
scattering process. If we assume that at some stage before hadronization 
the target field is separated from the rest of the final state, the 
inclusiveness of the process translates into a summation over all field 
configurations in the cross section. This corresponds exactly to the 
discussion of the previous section, where a summation over all field 
configurations of the target had to be performed for the cross section of 
DIS.

\section{Angular distributions}\label{dyad}
In DY pair production the transverse and longitudinal photon polarizations 
can be distinguished by measuring the angle between the direction of the 
decay lepton and the $z$-axis. However, more information can be obtained by 
considering the azimuthal angle as well. In particular, additional 
integrals involving the $q\bar{q}$-cross section $\sigma(\rho)$ are 
provided by the angular correlations.

As explained in the last section the $u$-channel frame is most suitable for 
an analysis along the lines of small-$x$ DIS. We work with a right-handed 
coordinate system, the $z$-axis being antiparallel to $\vec{p}_t$ and the 
$y$-axis parallel to $\vec{p}_p\times\vec{p}_t$, where $\vec{p}_p$ and 
$\vec{p}_t$ are the projectile and target momenta in the photon rest frame 
(see e.g. \cite{fal}). 

The direction of the produced lepton is characterized by the standard 
polar and azimuthal angles $\theta$ and $\phi$. To obtain the complete 
angular dependence of the cross section, interference terms between 
different photon polarizations have to be considered in equations analogous 
to (\ref{dycs}) and (\ref{dyt}). The obtained contributions are multiplied 
by typical angle dependent functions obtained from the leptonic tensor (for 
details see e.g. \cite{mir}). 

In general, the angular dependence can be given in the form
\be
\frac{1}{\sigma}\frac{d\sigma}{d\Omega}\sim1+\lambda\cos^2\theta+\mu\sin2
\theta\cos\phi+\frac{\nu}{2}\sin^2\theta\cos2\phi\,.
\ee
The cross section will be presented after integration over the transverse 
momentum of the pair. The dependence on $q_\perp^2$ can be recovered from 
the formulae in the appendix, where some details of the calculation are 
given. 

In compact notation the results of our impact parameter space calculation 
of the DY cross section read
\be
\frac{d\sigma}{dx_FdM^2d\Omega}=\frac{\alpha_{\mbox{\scriptsize em}}^2}
{2(2\pi)^4M^2}\!\!\!\!\!\!\!\int\limits_0^{(1-x_F)/x_F}\!\!\!\!\!\!\!d
\alpha\int d^2r_\perp\frac{q\Big(x_F(1+\alpha)\Big)}{(1+\alpha)^2}
\sigma(r/N)\sum_if_i(\alpha,r)h_i(\theta,\phi)\, ,\label{dya}
\ee
where $i\in \{T,\,L,\,TT,\,LT\}$ labels the contributions of transverse and 
longitudinal polarizations and of the transverse-transverse and 
longitudinal-transverse interference terms. Note, that in contrast to 
Eq.~(\ref{dyex}) here the integration is over the dimensionless variable 
$r_\perp=N\rho_\perp$. The angular dependence is given by the functions 
\begin{eqnarray}
h_T(\theta,\phi)=1+\cos^2\theta&,&\qquad h_{TT}(\theta,\phi)=
\sin^2\theta\cos2\phi\, ,\label{h1}\\
h_L(\theta,\phi)=1-\cos^2\theta&,&\qquad h_{LT}(\theta,\phi)=\sin2\theta
\cos\phi\,.\label{h2}
\end{eqnarray}
Finally, the $\alpha$- and $r$-dependent coefficients read
\begin{eqnarray}
f_T(\alpha,r)&=&[\alpha^2+(1+\alpha)^2]K_1^2(r)\\
f_L(\alpha,r)&=&4\alpha(1+\alpha)K_0^2(r)\\
f_{TT}(\alpha,r)&=&\alpha(1+\alpha)\Big[r^{-1}K_1''(r)+r^{-2}K_1'(r)-r^{-3}
K_1(r)-2K_1^2(r)\Big]\\
f_{LT}(\alpha,r)&=&r^{-1}(1+2\alpha)\sqrt{\alpha(1+\alpha)}\Big[K_0(r)\Big(
rA(r)-1\Big)\label{flt}\\
&&-K_1(r)\Big((2r)^{-1}-rA'(r)\Big)-K_1'(r)/2\Big]\, .\nonumber
\end{eqnarray}
Here the first two functions give the transverse and longitudinal 
contributions of the last section. The function $A$ is defined by the 
following definite integral, that can be expressed through the difference 
of the modified Bessel function $I_0$ and the modified Struve function 
$\mbox{\bf L}_0$ \cite{gr}, 
\be
A(r)=\int_0^\infty\frac{dt\sin rt}{\sqrt{1+t^2}}=\frac{\pi}{2}\Big(I_0(r)-
\mbox{\bf L}_0(r)\Big)\, .
\ee

As discussed in the previous section the integral involving $f_T$ receives 
a contribution from large $\rho$. In Eq.~(\ref{dya}) this is most easily 
seen by recalling that $\sigma(\rho)\sim\rho^2$ at small $\rho$. Replacing 
$\sigma(r/N)$ with the model form $r^2/N^2$ results in a divergent 
$\alpha$-integration. This shows the sensitivity to the large 
$\rho$-behavior of $\sigma(\rho)$. In DY pair production on nuclei this 
sensitivity will show up as leading-twist shadowing, since configurations 
with large cross section are absorbed at the surface. This is analogous to 
the leading-twist shadowing in DIS \cite{fs}. 

In contrast to the integral of $f_T$, the integrals involving $f_L,\, 
f_{TT}$ and $f_{LT}$ are dominated by the region of small $\rho$ at leading 
twist. To see this, notice that replacing $\sigma(r/N)$ with $r^2/N^2$ in 
Eq.~(\ref{dya}) results in finite $\alpha$-integrations for $f_L,\, f_{TT}$ 
and $f_{LT}$. This leading-twist contribution corresponds to the effect of 
the gluon distribution of the target. Integrations involving higher powers 
of $r$ are sensitive to large transverse distances, but they are suppressed 
by powers of $M$. This corresponds to the fact that in the leading order 
(and leading-twist) parton model these angular coefficients vanish. 

The above discussion shows that in the longitudinal contribution and in the 
interference terms, shadowing appears only at higher twist or at higher 
order in $\alpha_S$. While higher-twist terms are suppressed by 
$\Lambda^2/M^2$ in the longitudinal cross section and in the 
transverse-transverse interference term, they are only suppressed by 
$\Lambda/M$ in the longitudinal-transverse interference. This results from 
the weaker suppression of $f_{LT}$ at small $\alpha$ (see Eq.~(\ref{flt})). 

The presented formulae contain contributions from all transverse sizes of 
the effective $q\bar{q}$-pair interacting with the target gluonic field, 
thus including all higher-twist corrections from this particular source. 
Our analysis also gives a simple and intuitive derivation of the 
dominant QCD-corrections at small $x$, associated with the gluon 
distribution of the target.

\section{Conclusions}\label{con}
A detailed calculation of the DY cross section, including its angular 
dependence, has been performed in the target rest frame in the limit of 
high energies and small $x_{\mbox{\scriptsize{target}}}$. The close 
similarity with the impact parameter description of DIS has been 
established for transverse and longitudinal photon polarizations and the 
availability of additional angular observables in the DY process has been 
demonstrated. 

As is well known, in the small-$x$ limit DIS can be calculated from the 
elastic scattering of the quark-antiquark component of the virtual photon 
wave function off the hadronic target. The DIS cross section is given by a 
convolution of the photon wave function with the $q\bar{q}$-cross section 
$\sigma(\rho)$. This picture holds even when the interaction of each of the 
quarks with the target is completely non-perturbative.

The cross section for DY pair production via transversely and 
longitudinally polarized massive photons can be given as a convolution of 
the above $q\bar{q}$-cross section with analogues of the transverse and 
longitudinal photon wave functions. These functions depend on the photon 
momentum fractions carried by the quarks and on the photon virtuality in 
exactly the same way as in DIS. For this analogy to hold polarization by 
polarization the DY process has to be analyzed in the $u$-channel frame. 
This frame corresponds to the $\gamma^*p$-frame of small-$x$ DIS, since it 
uses photon and target hadron momenta for the definition of the $z$-axis. 

As in the DIS case, the transverse photon contribution is sensitive to 
large distances in impact parameter space. It receives a leading-twist 
contribution from large $\rho$, which corresponds to the effect of a 
non-perturbative antiquark distribution in the target. Our approach 
includes, beyond this leading-twist contribution and the 
$\alpha_S$-correction from small $\rho$, all higher-twist terms associated 
with different transverse distances inside the target. The universal 
function $\sigma(\rho)$ relates these contributions directly to the 
analogous terms in DIS. 

In addition to the transverse-longitudinal analysis, which can be performed 
using the polar angle of the produced lepton, the azimuthal angle allows the 
investigation of interference terms of different polarizations. These terms 
involve convolutions of $\sigma(\rho)$ with new functions, not available in 
DIS. 

Our analysis shows that rather detailed information about the function 
$\sigma(\rho)$ can be obtained from a sufficiently precise measurement of 
angular correlations in the DY process at small 
$x_{\mbox{\scriptsize{target}}}$. Even more could be learned from a 
measurement of the nuclear dependence of these angular correlations. Using 
the Glauber approach to nuclear shadowing, this type of measurement would 
provide additional information about the functional dependence of the 
$q\bar{q}$-cross section on $\rho$. We expect that future measurements of 
the DY process will help to disentangle the interplay of small and large 
transverse distances in small-$x$ physics. 

Several aspects of the presented approach require further study: 

The leading-twist part of our calculation combines the standard 
$q\bar{q}$-annihilation cross section with the $\alpha_S$-corrections 
associated with the target gluon density. It is certainly necessary to 
include other $\alpha_S$-corrections systematically into our approach. For 
example, corrections associated with the radiation of a gluon off the 
projectile quark can be treated in the impact parameter space by methods 
developed in \cite{bdh}.

Furthermore, higher-twist contributions from sources not considered here 
should be carefully analyzed. At small $x_{\mbox{\scriptsize{target}}}$, 
corresponding to large $x_F$, higher-twist corrections from comoving 
projectile partons are potentially important \cite{bb}.

Finally, to go beyond the classical field model, we have argued that the 
summation over all field configurations of the target is identical in DIS 
and the DY process. It would be highly desirable, to derive this statement 
in the framework of QCD and to specify the type of expected corrections.
\\*[0cm]

We would like to thank M.~Beneke, W.~Buchm\"uller, L.~Frankfurt, P.~Hoyer, 
B.~Kopeliovich, A.H.~Mueller, M.~Strikman, and R.~Venugopalan for valuable 
discussions and comments. A.H. and E.Q. have been supported by the Feodor 
Lynen Program of the Alexander von Humboldt Foundation.

\section*{Appendix}
Some details of the calculations leading to the results of Sections 
\ref{dy} and \ref{dyad} are presented below.

In analogy to Eq.~(\ref{dycs}) the angular distribution of the lepton in 
the DY process is given by
\be
\frac{d\hat{\sigma}}{dx_FdM^2d\Omega}=\frac{e^2}{192(2\pi)^4}\cdot\frac{1}
{x_Fk_0k_0'M^4}\int\frac{d^2q_\perp}{(2\pi)^2}\frac{d^2k_\perp'}{(2\pi)^2}
\Big(T_\lambda T^*_{\lambda'}\Big)\,L^{\mu\nu}\epsilon^\lambda_\mu
\epsilon^{\lambda'*}_\nu\, ,\label{adycs}
\ee
where the polarization sum is understood. The leptonic tensor $L^{\mu\nu}$ 
is contracted with the photon polarization vectors
\be
\epsilon_\pm=(0,1,\pm i,0)\, ,\quad \epsilon_0=(0,0,0,1)\, ,\label{defe}
\ee
defined in the $u$-channel frame, which has been specified in 
Sect.~\ref{dyad}. This expression gives the functions $h_i(\theta,\phi)$, 
introduced in Eqs.~(\ref{h1}),(\ref{h2}).

The amplitudes $T_\lambda$ are most conveniently calculated in the target 
rest frame, in a system where $q_\perp=0$. This corresponds to the 
$u$-channel frame, boosted appropriately along its $z$-axis. Of course now 
the amplitude is a function of $k_\perp$ $(k_\perp\neq0)$, and the 
$q_\perp$-integration in Eq.~(\ref{adycs}) has to be replaced by a 
$k_\perp$-integration,
\be
\int\frac{d^2q_\perp}{q_0^2} \rightarrow \int\frac{d^2k_\perp^2}{k_0^2}\, ,
\ee
leading to
\begin{eqnarray}
\!\!\!\!\!
\frac{d\hat{\sigma}}{dx_FdM^2d\Omega}&=&\frac{e^2}{96(2\pi)^4}\cdot
\frac{q_0^2}{x_Fk_0^3k_0'M^4}\int\frac{d^2k_\perp d^2k_\perp'}{(2\pi)^4}
\Bigg[\frac{h_T}{2}\Big(|T_+|^2+|T_-|^2\Big)+h_L|T_0|^2\nonumber\\
\!\!\!\!\!\label{csa}\\
\!\!\!\!\!
&&-\frac{h_{TT}}{2}\Big(T_+T_-^*+T_-T_+^*\Big)-\frac{h_{LT}}{2\sqrt{2}}
\Big(T_0T_+^*+T_0T_-^*+T_+T_0^*+T_-T_0^*\Big)\Bigg]\, .\nonumber
\end{eqnarray}

Now the amplitudes have to be calculated explicitly. In the high energy 
approximation the fermion propagators appearing in Eq.~(\ref{dyt}) can be 
treated as follows, 
\be
\frac{1}{\ppl}\approx\frac{\sum_r u_r(p)\bar{u}_r(p)}{p^2}\, .
\ee
Here $u(p)\equiv u(p_+,\bar{p}_-,p_\perp)$, with $\bar{p}_-\equiv 
p_\perp^2/p_+$, for off-shell momentum $p$. This approximation, which has 
been used in the above form in \cite{bdh}, corresponds to dropping the 
instantaneous terms in light-cone quantization \cite{bks}. 

In our frame with $q_\perp=0$ the resulting spinor products in the high 
energy approximation are given by 
\be
\bar{u}(k-q)\epsilons_\lambda u(k)\equiv g_\lambda(k_\perp,q_0,\alpha)\,
,\quad\bar{u}(k')\epsilons_\lambda u(k'+q)\equiv g_\lambda(k_\perp',q_0,
\alpha)\, ,\label{gg}
\ee
where the helicity dependence has been suppressed. By choosing some spinor 
representation explicit formulae are easily obtained. 

Using the definition of $\tilde{t}_q$ in Eq.~(\ref{tdef}), the following 
expression for the product of two amplitudes can now be given, 
\be
T_\lambda T_{\lambda'}^*=\Big[2eq_0\alpha(1\!+\!\alpha)\Big]^2\,\Big|
\tilde{t}_q(k_\perp'\!-\!k_\perp)\Big|^2\,\left(\frac{g_\lambda(k_\perp)}
{k_\perp^2\!+\!N^2}-\frac{g_\lambda(k_\perp')}{k_\perp'^2\!+\!N^2}\right)
\left(\frac{g_{\lambda'}(k_\perp)}{k_\perp^2\!+\!N^2}-\frac{g_{\lambda'}
(k_\perp')}{k_\perp'^2\!+\!N^2}\right)^*.\label{tt}
\ee
Here the $q_0$- and $\alpha$-dependence of the function $g$ has been 
suppressed. To relate this formula to the $q\bar{q}$-cross section of 
Sect.~\ref{dis}, observe that in the high energy limit the quark and 
antiquark scattering amplitudes are dominated by gluon exchange. This 
implies the relation $t_{\bar{q}}(x_\perp)=-t^\dagger_q(x_\perp)$. Note in 
particular, that this relation is respected by the eikonal approximation 
Eq.~(\ref{eik}). From the definition of $\sigma(\rho)$ in Eq.~(5) one now 
derives 
\be
|\tilde{t}_q(k_\perp'\!-\!k_\perp)|^2=-\frac{3}{2}\int d^2\rho_\perp
\sigma(\rho)\,e^{i\rho_\perp(k'-k)_\perp}+2(2\pi)^2\delta^2(k_\perp'\!-\!
k_\perp)\,\mbox{Im}\,\tilde{t}_q(0)\, .\label{ts}
\ee
When inserted into Eq.~(\ref{tt}) the second term on the r.h.s. of 
Eq.~(\ref{ts}) does not contribute, so that the product $T_\lambda 
T_{\lambda'}^*$ can indeed be expressed through the the $q\bar{q}$-cross 
section $\sigma(\rho)$. 

Two remarks have to be made concerning the treatment of the photon 
polarization vectors in Eqs.~(\ref{gg}),(\ref{tt}): 

First, recall that Eq.~(\ref{tt}) holds in the target rest frame with 
$z$-axis parallel to $\vec{q}$. Therefore, the transverse polarizations are 
the same as in the $u$-channel frame (see Eq.~(\ref{defe})). However, 
before the $k_\perp$-integration in Eq.~(\ref{csa}) can be performed, the 
$k_\perp$-dependence of the orientation of $x$- and $y$-axis has to be 
explicitly introduced. This is most easily done by assuming $k$ to be 
exactly parallel to the projectile momentum and writing $\hat{e}_x=k_\perp/ 
|k_\perp|$ (see the definition of the $u$-channel frame in 
Sect.~\ref{dyad}). 

Second, the boost from the $u$-channel frame to the target rest frame 
transforms the longitudinal polarization vector to 
\be
\epsilon_0=\frac{q}{M}-\frac{M}{q_0}(1,\vec{0})\, .
\ee
However, taking advantage of gauge invariance, the first term can be 
dropped. This significantly simplifies the evaluation of the corresponding 
spinor products in Eq.~(\ref{gg}). 

It is now straightforward to choose an explicit spinor representation, to 
evaluate Eq.~(\ref{gg}), and to combine Eqs.~(\ref{ts}),(\ref{tt}) and 
(\ref{csa}). Performing the $k_\perp$ and $k_\perp'$-integrations, the 
result stated in Eq.~(\ref{dya}) is obtained.

\end{document}